\documentclass[preprint,letterpaper,12pt]{JHEP3}
\usepackage{axodraw}
{\setlength\paperheight {11in} 
\setlength{\topmargin}{1in}

\newcommand{\GeV}{\,{\rm GeV}}

\def\bea{\begin{eqnarray}}
\def\eea{\end{eqnarray}}
\def\ba{\begin{array}}
\def\ea{\end{array}}
\def\bec{\begin{center}}
\def\ec{\end{center}}
\def\nn{\nonumber}
\def\la{\langle}
\def\ra{\rangle}

\def\ps{ SU(4)_C \times SU(2)_L \times SU(2)_R}

\def\64{\rm SO(6) \times SO(4)}

\def\f{\frac}
\def\f#1#2{\frac{#1}{#2}}

\setcounter{page}{1}
\preprint{OHSTPY-HEP-T-02-014\\KIAS-P02075}
\title{\huge Unification in 5D SO(10)}
\author{Hyung Do Kim$^{a,b}$ and Stuart Raby$^a$\\
$^a$Department of Physics, The Ohio State University,\\
174 W. 18th Ave., Columbus, Ohio 43210, USA\\
$^b$School of Physics,Korea Institute for Advanced Study,
Seoul, 235-010, Korea
\\ E-mail: \email{hdkim,raby@mps.ohio-state.edu}}
\abstract{
Gauge unification in a five dimensional supersymmetric $SO(10)$
model compactified on an orbifold $S^1/(Z_2 \times Z_2^{\prime})$
is studied. One orbifolding reduces $N=2$ supersymmetry to $N=1$,
and the other breaks $SO(10)$ to the Pati-Salam gauge group $\ps$.
Further breaking to the standard model gauge group is made through
the Higgs mechanism on one of the branes. The differences of the
three gauge couplings run logarithmically even in five dimensions
and we can keep the predictability for unification as in four
dimensional gauge theories.  We obtain an excellent prediction for
gauge coupling unification with a cutoff scale $M_* \sim 3 \times
10^{17}$ GeV and a compactification scale $M_c \sim 1.5 \times
10^{14}$ GeV.   Finally, although proton decay due to dimension 5
operators may be completely eliminated,  the proton decay rate in
these models is sensitive to the placement of matter multiplets in
the 5th dimension, as well as to the unknown physics above the
cutoff scale.
}
\begin{document}

\section{Introduction}

Unification of the forces of nature is one of the most important
ideas in the fundamental physics of the twentieth century.  Early
attempts at a unified field theory, including gravity and
electromagnetic interactions, were in hindsight premature. However
the Standard Model, describing with remarkable success the strong,
weak and electromagnetic interactions of the three families of
quarks and leptons, seems to provide a much more propitious
starting point for unification.   It is entirely possible that
there are no new low energy gauge interactions. Moreover quarks
and leptons appear to be equally fundamental and elementary.
Finally, the merging of the three low energy gauge couplings at a
high energy scale $M_G = 3 \times 10^{16}$ GeV provides
significant evidence for a grand unification within four
dimensional supersymmetric [SUSY]
models\cite{Dimopoulos:1981yj,Dimopoulos:1981zb,Amaldi:1991cn}.

Let us consider some other virtues of 4D SUSY GUTs and also some
of the problems.  On the plus side:
\begin{itemize}
\item GUTs explain the charge assignments of a single family of
quarks and leptons, and the quantization of
$U(1)_Y$\cite{Pati:uk,Georgi:sy,Georgi:my}. Recall that in $SU(5)$
the quarks and leptons of one family are described by $ \{
Q_{+1/6} = \left(\begin{array}{c} u \\ d
\end{array}\right), \;\;\; \bar {e}_{+1}, \;\;\; \bar{u}_{-2/3}
\} \;\; \subset \;\; \bf 10$ and
$ \{ \bar{d}_{+1/3}, \;\;\; L_{-1/2} = \left(\begin{array}{c} \nu \\
e \end{array}\right) \} \;\; \subset \;\; \bf \bar{5}$ (where the
subscript is the hypercharge $Y$ with electric charge given by $Q
= T_{3 L} + Y$).   The two Higgs doublets of the minimal
supersymmetric standard model [MSSM] are given by $ (H_u)_{+1/2}
\;\; \subset \;\; {\bf 5_H}$ and $(H_d)_{-1/2} \;\; \subset \;\;
{\bf \bar{5}_H}$. In $SO(10)$ we have the more compelling
unification of all quarks and leptons of one family into one
irreducible spinor representation with  $ \{ {\bf 10}, \;\; {\bf
\bar{5}}, \;\; \bar{\nu}_{sterile} \} \;\; \subset \;\; { \bf 16}$
and the two Higgs doublets are also unified with $ \{ {\bf
5_H},\;\; {\bf \bar{5}_H} \} \;\; \subset \;\; {\bf 10_H} $.

\item The sterile neutrino is a natural ingredient of the $SO(10)$
spinor representation.   Hence, the observed oscillations of
neutrinos are nicely explained by incorporating the See-Saw
mechanism; giving an extremely light mass to the active neutrinos
when the sterile neutrino mass is of order $
M_G$\cite{Gell-Mann:vs,yanagida}.

\item In $SU(5)$, the bottom quark and tau lepton Yukawa couplings
are predicted to be equal at the unification scale. This is
consistent with the observed data for small $\tan\beta \sim 1$ or
large $\tan\beta \sim 50$\cite{Barr:2002mw}.   However, LEP data
favors values of $\tan\beta > 2.4$\cite{Heister:2001kr}. Hence
$\lambda_b = \lambda_\tau$ Yukawa unification works best for large
$\tan\beta \sim 50$ which is also consistent with $SO(10)$ Yukawa
unification with $\lambda_t = \lambda_b = \lambda_\tau =
\lambda_{\bar \nu_\tau}$\cite{Blazek:2001sb}.

\end{itemize}

On the other hand, the concrete realization of the unification
idea generates several problems.
\begin{itemize}
\item Doublet-Triplet Higgs splitting: Though matter fields make
complete multiplets under the unified gauge group, it is not true
for the Higgs fields which are essential for giving mass at the
electroweak scale to the W and Z bosons, quarks and leptons.
Inevitably colored Higgs fields are needed to form complete GUT
multiplets. Moreover, they must be extremely heavy in order not to
make the proton decay too fast. Understanding why Higgs doublets
remain light while Higgs triplets become heavy is one of the most
difficult problems in any unified theory. There are, in fact,
several proposed solutions to the problem, such as the missing
partner or missing vacuum expectation value [vev]
mechanism\cite{Dimopoulos:zu}. But none of these are sufficiently
attractive to be accepted universally as the solution.

\item Proton decay has yet to be observed.  This is not a problem
for dimension 6 operators which predict a proton lifetime of order
$5 \times 10^{36 \pm 1}$ yrs\cite{Murayama:2001ur}, but seriously
constrains dimension 5 operators which, with reasonable
assumptions, has an upper bound of $\tau( p \rightarrow K^+ \bar
\nu ) \leq (1/3 - 3) \times 10^{33}$ yrs\cite{Babu:1998wi}.

\item Minimal supersymmetric $SU(5)$ gauge symmetry breaking is
ruled out by the non-observation of proton decay
\cite{Goto:1998qg,Murayama:2001ur}.   However, more complicated
GUT breaking sectors, in $SU(5)$ and $SO(10)$, with natural D-T
splitting are acceptable \cite{Babu:1998wi}.

\item Yukawa unification works for the third generation, but not
for the first and second generations, since $m_s/m_d \neq
m_\mu/m_e$.  Therefore, understanding the flavor structure of
fermion masses and mixing remains a separate problem.  However,
predictive and realistic theories of fermion masses can be
obtained using family symmetries to relate quark and lepton mass
matrices\cite{Barbieri:1996ww,Blazek:1999ue}
\end{itemize}

Recent developments in theories with extra dimensions allow us to
rethink these traditional problems from a different perspective.
$SU(5)$ models on $S^1/(Z_2 \times Z_2^{\prime})$ have been
studied (for example, see
\cite{Kawamura:2000ev}-\cite{Hebecker:2001wq}). It has been shown
that $SU(5)$ orbifold GUTs [OGUTs] have simple predictions for
gauge coupling unification when a strong coupling assumption is
used. $SO(10)$ OGUTs have also been studied in both five
\cite{Dermisek:2001hp}-\cite{Kyae:2002ss} and six dimensions
\cite{Asaka:2001eh}. Two extra dimensions have been introduced to
utilize two steps of symmetry breaking from $SO(10)$ to the
standard model gauge group. However the usual orbifolding
procedure cannot reduce the rank in this
process.\footnote{Although it is known to be possible to reduce
the rank of a gauge group by orbifolding (using non-abelian
discrete symmetries\cite{Hall:2001tn} or charge
conjugation\cite{Hebecker:2001jb}), neither of these approaches
work when breaking $SO(10)$ to the Standard Model.} As a result
one additional $U(1)$ always remains and must be broken
spontaneously via the traditional Higgs mechanism. Accepting this
fact, one extra dimension is rather economical or minimal.

The addition of one extra dimension (defined as a finite line
segment) has some advantages over conventional four dimensional
GUTs. \begin{itemize} \item GUT symmetry breaking, without a
complicated Higgs sector, can be accomplished via Wilson line
breaking or equivalently, an orbifold parity.\footnote{In most
cases studied in the literature, such a symmetry breaking vev is
an ad hoc assumption. Indeed, dynamical arguments for why such a
vacuum is preferred have been considered (see
\cite{Haba:2002py}).}

\item  In addition, doublet-triplet Higgs splitting is naturally
explained by different orbifold parities. The triplet Higgs has a
twisted boundary condition while the doublet Higgs has an
untwisted one. Thus the mass of the triplet Higgs is inversely
proportional to the size of the extra dimension. This naturally
explains why only Higgs doublets remain light. Indeed this is the
bottom-up version of the string theory compactification
extensively studied in the 80's\cite{Dixon:jw}.

\item Also, the prediction for gauge coupling unification is, with
some reasonable assumptions, just as predictive as in four
dimensional SUSY GUTs.  Threshold corrections due to the Higgs and
GUT breaking sectors of the four dimensional theories are replaced
by towers of Kaluza-Klein modes of gauge and Higgs fields.
Although higher dimensional theories are non-renormalizeable,
nevertheless the orbifold breaking of the gauge symmetry still
gives a logarithmic running for the differences of the gauge
couplings.  This makes it possible to have a prediction for gauge
coupling unification as clean as that of four dimensional GUTs.

\item Finally, the ``dangerous" baryon and lepton number violating
dimension 5 operators can be forbidden by R symmetries (see, for
example, Hall and Nomura \cite{Hall:2001pg}).
\end{itemize}

Though higher dimensional GUTs solve some of the problems of four
dimensions ones, at the same time they present new problems.
\begin{itemize} \item A higher dimensional gauge theory is
nonrenormalizeable and must be defined as a cutoff theory. This
implies a highly sensitive dependence on unknown short distance
physics.  This is a problem for gauge coupling unification if it
affects how the difference of any two gauge couplings run above
the compactification scale.

Let us elaborate on the last statement (we restrict our discussion
to one extra dimension).  Regarding unification, the differences
of gauge couplings are less sensitive to the short distance
physics in certain cases. If the gauge symmetry is broken by
orbifolding or via Higgs vevs on the brane (i.e. the fixed point),
the difference of three gauge couplings still gets only a
logarithmic correction though the individual couplings get power
law corrections.\footnote{Though we restrict our consideration to
the flat extra dimensions in this paper, there are other ways of
obtaining logarithmic running in $AdS_5$. When there is a gauge
theory on a curved slice of $AdS_5$, the relation between the
unified gauge coupling and the gauge couplings at the electroweak
scale involves only logarithmic corrections as long as the
symmetry breaking scale is low compared to the curvature scale of
$AdS_5$. This possibility has received some attention recently
\cite{Pomarol:2000hp}. Orbifold breaking of gauge symmetry in
$AdS_5$ is another way \cite{Choi:2002wx}.} On the contrary if the
gauge symmetry is broken by the vacuum expectation value of a bulk
field, the differences of the gauge couplings are also sensitive
to the physics of the cutoff scale. In this latter case,
predictability is generically lost.  However in 5D, imposition of
a $Z_2$ symmetry may preserve predictability (see Ref.
\cite{Hebecker:2002vm}).

\item Another problem results from orbifolding. For 5D $SU(5)$,
the simple orbifold symmetry breaking gives a symmetric fixed
point at which $SU(5)$ is realized, as well as a Standard Model
[SM] fixed point on which only the SM gauge symmetry survives. As
a result, the GUT virtues of explaining charge quantization or
explaining the SM charge assignment of a single family of quarks
and leptons is lost.

\begin{itemize}
\item Let us elaborate on this last comment.  Charge quantization
is lost since arbitrary $U(1)$ charges are allowed for fields
living at the SM fixed point.

\item Only the third generation can reside on the symmetric
$SU(5)$ brane due to constraints of proton decay (see below).
Hence, as a consequence of orbifold GUT symmetry breaking, a first
or second generation SM family is necessarily a collection of
states (free from SM gauge anomalies) coming from different
$SU(5)$ multiplets in the bulk or states from the SM brane.   Of
course, this is no better than in the SM.

\item In addition, GUT mass relations are lost unless the families
live on the symmetric fixed point.  Thus, it does not allow for
the successful Georgi-Jarlskog relation\cite{Georgi:1979df}) for
the first and second generations obtained in predictive
theories\cite{Barbieri:1996ww,Blazek:1999ue}.
\end{itemize}
\item Finally, baryon and lepton number violating dimension 6
operators, on the $SU(5)$ symmetric brane, are only suppressed by
the compactification scale $O(10^{14 - 15}) \; {\rm GeV}$, which
is significantly less than the 4D GUT scale.
\end{itemize}

These problems are resolved in a 5D $SO(10)$ SUSY GUT. Consider
the model given first in \cite{Dermisek:2001hp}.  In this model,
$SO(10)$ is broken down to Pati-Salam $\ps$ by a $Z_2 \times
Z_2^{\prime}$ orbifold twisting which generates two inequivalent
fixed points. One is an $SO(10)$ symmetric fixed point [we call
this the $SO(10)$ brane] while the other has only a Pati-Salam
symmetry [we refer to as the PS brane]. Unlike the 5D $SU(5)$
model, or 6D $SO(10)$ models, the broken gauge group at the fixed
points does not contain $U(1)$ factors. Further breaking to the
standard model can be done by making $\chi^c + \bar \chi^c$
($\chi^c = {\bf (\bar 4, 1, \bar 2))}$ acquire vevs on the brane.
We restrict our Higgs breaking to the brane in order to maintain
predictability for gauge coupling unification. Thus this model has
the following virtues:
\begin{itemize} \item There is no $U(1)$ factor in the remaining
gauge group after orbifold breaking  -- hence an explanation of
charge quantization for quark and lepton families is
retained.\footnote{Note, it is still possible to have states with
exotic SM charges, even though we have charge quantization.
 For example, vector-like PS representations, such as ${\bf (4,1,1) + (\bar 4,1,1)}$,
have exotic SM charges.   These states are allowed on the PS
brane.  However, they are expected to have mass of order the
cutoff scale, since a mass term is not forbidden by any symmetry.
Such states, in fact, are obtained in string
compactifications\cite{Leontaris:1995td}}

\item We preserve an $SU(4)_C \times SU(2)_L \times SU(2)_R$
Pati-Salam symmetry after orbifolding.  Quarks and leptons of one
SM family are contained in two irreducible representations, $\psi
\equiv \{ Q, \; L \} \;\; \subset \;\; {\bf (4, 2, 1)}$ and $
\psi^c \equiv \{ \left( \begin{array}{c} \bar u \\ \bar d
\end{array}\right) , \; \left( \begin{array}{c} \bar \nu \\ \bar e
\end{array}\right)  \} \;\; \subset \;\;
{\bf (\bar 4, 1, \bar 2)}$.

Moreover, this symmetry is sufficient for obtaining most fermion
mass relations of $SO(10)$. For example, third generation Yukawa
unification is a consequence of PS, if we also require the minimal
Higgs content, with both $ \{ H_u, \ H_d \} \equiv {\cal H} \;\; =
\;\; {\bf (1, \bar 2, 2)} \;\; \subset \;\; {\bf 10}$.  The
minimal Yukawa coupling is then given by  \bea \lambda \ \psi \
{\cal H} \ \psi^c . \eea The Higgs fields, as well as the matter
multiplets, may either be in the bulk or on the PS
brane.\footnote{As before, only the third family can reside on the
$SO(10)$ brane due to proton decay constraints.} Also, it is easy
to obtain Georgi-Jarlskog relations for the first and second
family on the PS brane using the B-L vev ($\propto
diag(1/3,1/3,1/3, -1)$) of an $SU(4)_C$ adjoint.

\item The final breaking to the Standard Model gauge group is done
via the standard Higgs mechanism on the brane [we refer to this as
brane breaking]. As a consequence the difference of gauge
couplings run logarithmically above the compactification scale.

\item Finally, proton decay, mediated by dimension 6 operators on
the $SO(10)$ brane, are suppressed by $1/M_c^2$.   This constrains
the matter multiplets which may live on this brane. However,  it
is possible to put all matter multiplets on the PS brane and still
retain fermion mass relations.    Indeed, with all matter on the
PS brane, dimension 6 proton decay is suppressed by the cutoff
scale $M_*$.  Unfortunately,  the proton lifetime is extremely
sensitive to unknown physics at the cutoff scale.   The decay $p
\rightarrow \pi^0 \ e^+$ may be observable but at the moment it
cannot be predicted.

\end{itemize}

In this paper, we review the 5D $SO(10)$ SUSY GUT construction of
Derm\' \i \v sek and Mafi\cite{Dermisek:2001hp}.  Our main result
is the detailed calculation of the Kaluza-Klein threshold
corrections to gauge coupling unification in this hybrid model
including both orbifold and Higgs symmetry breaking.  We find
excellent agreement with the low energy data with a cutoff scale
$M_* \sim 3 \times 10^{17}$ GeV and a compactification scale $M_c
\sim 1.5 \times 10^{14}$ GeV. The large ratio $M_*/M_c \sim 10^3$
is necessary to preserve gauge coupling unification in all higher
dimensional GUTs.   We argue that in our model a non-perturbative
UV fixed point may exist which makes such a large ratio plausible.
Finally, proton decay from dimension 5 operators may be eliminated
by an R symmetry. However the contribution of dimension 6
operators is highly sensitive to the placement of matter
multiplets in the extra dimension.  If all matter resides either
in the bulk or on the PS brane, the proton lifetime cannot be
predicted since it depends significantly on the physics at the
cutoff scale.

\section{Setup}

We consider 5D supersymmetric $SO(10)$ gauge theory compactified
on an $S^1/(Z_2 \times Z_2^{\prime})$ orbifold. $Z_2$ breaks half
of the supersymmetry and $Z_2^{\prime}$ breaks the gauge symmetry
$SO(10)$ down to Pati-Salam [PS] gauge group $\ps$. Then $\chi^c +
\bar \chi^c$ on the PS fixed point breaks PS down to the Standard
Model gauge group (see Figure 1).\footnote{We could also use a
${\bf \overline{16} + 16}$ on the $SO(10)$ symmetric brane to
break $SO(10)$ to $SU(5)$.  The unbroken gauge group in the
overlap of $\ps$ with $SU(5)$ is just the SM (see Figure 2).}

\begin{figure}
\begin{center}
\begin{picture}(400,250)(-20,50)
\GCirc(200,150){100}{1} \GCirc(200,130){80}{0.9}
\GCirc(200,110){60}{0.8} \Text(200,230)[]{$F$ = $SO(10)$}
\Text(200,190)[]{Orbifold breaking:} \Text(200,177)[]{$G$
=Pati-Salam} \Text(200,120)[]{Brane breaking:}
\Text(200,105)[]{$H$ = Standard Model}
\end{picture}
\caption{A diagram showing the gauge symmetry breaking when the
Higgs mechanism is on the Pati-Salam brane by $\chi^c$. The big
circle shows the largest gauge group F ($SO(10)$ in our specific
example), and the bright gray circle represents the unbroken
subgroup G after the breaking by the orbifolding (G = Pati-Salam
gauge group in our specific example). The dark gray circle is the
unbroken subgroup $H$  after the breaking by the VEV of the brane
fields, i.e., the usual Higgs mechanism (H = the Standard Model
gauge group in our specific example). Here $G^{\prime} = H \subset
G$. We consider this as our setup.}
\end{center}
\end{figure}
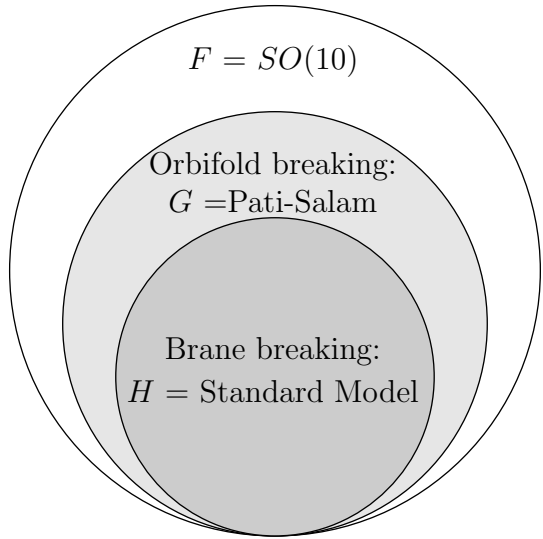

\begin{figure}
\begin{center}
\begin{picture}(400,250)(-20,50)
\GOval(150,170)(100,150)(0){1} \GCirc(185,150){70}{0.8}
\GCirc(115,150){70}{0.9} \CArc(185,150)(70,90,270)
\Text(150,240)[b]{$F$ = $SO(10)$} \Text(60,175)[l]{Orbifold}
\Text(60,160)[l]{breaking:} \Text(65,145)[l]{$G$ =}
\Text(55,130)[l]{Pati-Salam} \Text(190,175)[l]{Brane}
\Text(190,160)[l]{breaking:} \Text(195,145)[l]{$G^{\prime}$ =}
\Text(192,130)[l]{SU(5)} \Text(150,170)[]{$H$ =}
\Text(150,155)[]{Standard} \Text(150,140)[]{Model}
\end{picture}
\caption{A diagram showing the gauge symmetry breaking when the
Higgs mechanism is on the SO(10) preserving brane by $\bf 16$
\cite{Dermisek:2001hp}. The big oval shows the largest gauge group
F ($SO(10)$ in our specific example), and the bright gray circle
represents the unbroken subgroup G after the breaking by the
orbifolding (Pati-Salam gauge group in our specific example). The
dark gray circle is the unbroken subgroup $G^{\prime}$ after the
breaking by the VEV of the brane fields, i.e., the usual Higgs
mechanism ($SU(5)$ in our specific example). The intersection of
the two circles corresponds to the remaining subgroup $H$ which is
unbroken even after both the orbifold and brane breaking given by
$H = G \cap G^{\prime}$ (the Standard Model gauge group in our
specific example).}
\end{center}
\end{figure}
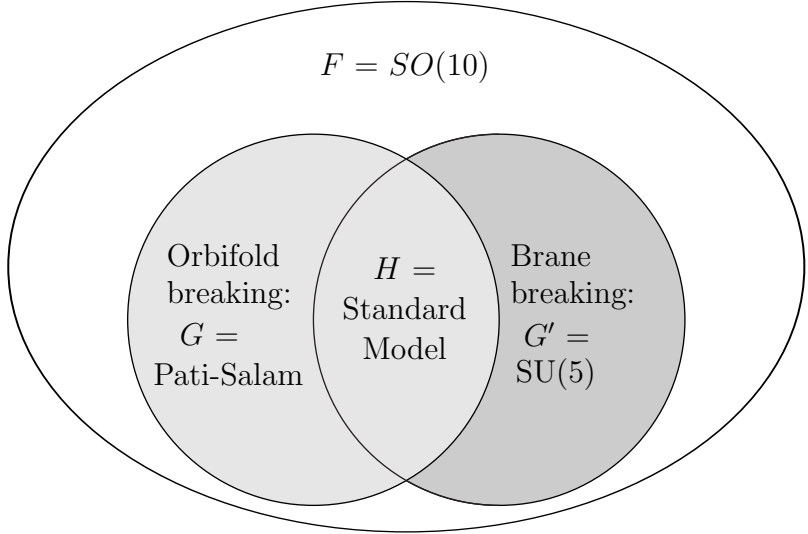

The 5D spacetime is a product of flat 4D spacetime $M^4$ with
$x^{\mu}$ ($\mu$=0,1,2,3) and an extra dimension compactified on
the $S^1/(Z_2 \times Z_2^{\prime})$ with y (=$x_5$) $= y + 2\pi
R$. $Z_2$ identifies $y \rightarrow -y$ which has a fixed point at
$y=0$ ($= \pi R$). $Z_2^{\prime}$ identifies $y - \pi R/2
\rightarrow -y + \pi R/2$ and the fixed point is at $y = \pi R/2 (
= 3\pi R/2)$. Combining $Z_2$ and $Z_2^{\prime}$, we get a
fundamental domain $y: [0, \pi R/2]$ and both ends are
inequivalent fixed points of $Z_2$ and $Z_2^{\prime}$,
respectively.

Under the orbifolding of $Z_2 \times Z_2^{\prime}$, a bulk field
$\phi(x^{\mu},y)$ transforms as follows
\bea
\phi(x^{\mu},y) & \rightarrow & \phi(x^{\mu},-y) = P \phi(x^{\mu},y), \\
\phi(x^{\mu},y-\pi R/2) & \rightarrow & \phi(x^{\mu},-(y-\pi R/2))
= P^{\prime} \phi(x^{\mu},y-\pi R/2), \eea where $P$ and
$P^{\prime}$ must have eigenvalues $\pm 1$. According to the
parity assignment, the fields have a Fourier mode expansion given
by \bea \phi_{++} (x^{\mu},y) & = & \sum_{n=0}^{\infty}
\frac{1}{\sqrt{2^{\delta_{n,0}}\pi R}}
\phi_{++}^{(2n)} (x^{\mu}) \cos \frac{2ny}{R},  \\
\phi_{+-} (x^{\mu},y) & = & \sum_{n=0}^{\infty} \frac{1}{\sqrt{\pi
R}}
\phi_{+-}^{(2n)} (x^{\mu}) \cos \frac{(2n+1)y}{R},  \\
\phi_{-+} (x^{\mu},y) & = & \sum_{n=0}^{\infty} \frac{1}{\sqrt{\pi
R}}
\phi_{-+}^{(2n)} (x^{\mu}) \sin \frac{(2n+1)y}{R}, \\
\phi_{--} (x^{\mu},y) & = & \sum_{n=0}^{\infty} \frac{1}{\sqrt{\pi
R}} \phi_{--}^{(2n)} (x^{\mu}) \sin \frac{(2n+2)y}{R},  \eea and
the lightest Kaluza-Klein mass of each field appears at 0, 1/R,
1/R, 2/R for $\phi_{++}$, $\phi_{+-}$, $\phi_{-+}$, $\phi_{--}$,
respectively.  From now on, $M_c = 1/R$ is used for the
compactification scale and all the Kaluza-Klein masses are integer
multiples of $M_c$.

\subsection{Supersymmetry breaking by $Z_2$}
5D N=1 supersymmetry has vector multiplets and hypermultiplets.
The 5D vector multiplet ${\cal V} =
({A_M,\lambda,\lambda^{\prime},\sigma})$ consists of one 4D vector
multiplet $V=(A_{\mu},\lambda)$ and one 4D chiral multiplet
$\Sigma = ((\sigma+iA_5)/{\sqrt{2}},\lambda^{\prime})$. The 5D
hypermultiplet ${\cal H} = (\phi, \phi^c, \psi, \psi^c)$ consists
of two 4D chiral multiplets $\Phi = (\phi,\psi)$ and $\Phi^c =
(\phi^c,\psi^c)$ which are in complex conjugate representations of
each other. A 5D N=1 multiplet is equivalent to a 4D N=2
multiplet.

Under the parity transformation $y \rightarrow -y$,
\bea
\left(\ba{l} V \nn \\ \Sigma \ea \right)
& \rightarrow & \eta_V
\left(\ba{l} +V \nn \\ -\Sigma \ea \right),  \\
\left(\ba{l} \Phi \nn \\ \Phi^c \ea \right) & \rightarrow &
\eta_{\Phi} \left(\ba{l} +\Phi \nn \\ -\Phi^c \ea \right), \eea
where $\eta$ can be $\pm$ depending on the individual multiplets.
For the 5D vector multiplet, the relative parity difference is
derived from the fact that $A_{\mu}$ and $A_5$ must transform
oppositely under the symmetry $y \rightarrow -y$. For the 5D
hypermultiplet, it can be read off from the bulk Lagrangian \bea W
\supset \Phi \partial_5 \Phi^c, \nn \eea which should be even
under $Z_2$. This relative difference of parity also applies to
$Z_2^{\prime}$. The zero modes consist of $(V,\Phi)$ for $\eta_V =
\eta_{\Phi} =1$. Therefore, 5D N=1 supersymmetry (or 4D N=2
supersymmetry) is broken to 4D N=1 supersymmetry by $Z_2$
orbifolding. {\footnote{More precisely the theory has more
symmetry given by the nontrivial relations between higher
Kaluza-Klein modes.}} Were we to compactify the 5D theory on a
circle we would obtain a vector-like theory since the 5D spinor
contains both the 4D left and right-handed spinors. ($\gamma_5$ is
the 5D gamma matrix.) Only the chiral projection $y \rightarrow
-y$ can eliminate one of them since $P\gamma_5 = -\gamma_5$.
Therefore we obtain a 4D chiral theory from a 5D theory by $Z_2$
orbifolding.

\subsection{Gauge symmetry breaking by $Z_2^{\prime}$}

Whereas $Z_2$ acts trivially on the internal space,
$Z_2^{\prime}$, on the other hand, acts nontrivially in order to
break the $SO(10)$ gauge symmetry at low energies. Following
Dermisek and Mafi, we use $P^{\prime} = {\rm diag} (-1,-1,-1,1,1)
\times {\rm diag} (1,1)$.\footnote{Note, $P^{\prime} \equiv
\exp({- i \frac{3 \pi}{2} \ (B - L)})$.  On spinor representations
it has eigenvalues $\pm \ i$.  Hence, on spinors $Z_2^\prime$ is
generated by $P^\prime = \exp{(- i \frac{3 \pi}{2} \ (B - L)})
\times P_F$ where $P_F$ is a discrete flavor symmetry with
eigenvalues $\pm \ i$, such that $(P^\prime)^2 = 1$.} Under the
$Z_2^{\prime}$ transformations, \bea V(x^{\mu},y-\pi R/2) &
\rightarrow & V(x^{\mu},-(y-\pi R/2))
= P^{\prime} V(x^{\mu},y-\pi R/2) P^{\prime} , \\
\Sigma (x^{\mu},y-\pi R/2) & \rightarrow & \Sigma (x^{\mu},-(y-\pi
R/2))
= -P^{\prime} \Sigma (x^{\mu},y-\pi R/2) P^{\prime} , \\
\Phi (x^{\mu},y-\pi R/2) & \rightarrow & \Phi (x^{\mu},-(y-\pi
R/2))
= P^{\prime} \Phi (x^{\mu},y-\pi R/2) , \\
\Phi^c (x^{\mu},y-\pi R/2) & \rightarrow & \Phi^c (x^{\mu},-(y-\pi
R/2)) = -P^{\prime} \Phi^c (x^{\mu},y-\pi R/2) . \eea

Therefore, the fields in different representations of the
Pati-Salam gauge group have different parities. According to the
representation, the 5D vector multiplet is decomposed as  (see
appendix A)
\bea V_{(15,1,1)++}, &&
\Sigma_{(15,1,1)--} \nn \\
V_{(1,3,1)++}, &&
\Sigma_{(1,3,1)--} \nn \\
V_{(1,1,3)++}, &&
\Sigma_{(1,1,3)--} \nn \\
V_{(6,2,2)+-}, && \Sigma_{(6,2,2)-+}, \nn \eea and only the
Pati-Salam vector multiplets have zero modes.
 Also the 5D
hypermultiplet is decomposed as \bea H_{(6,1,1)+-}, &&
H^{c}_{(6,1,1)-+} \nn \\
H_{(1,2,2)++}, && H^{c}_{(1,2,2)--}, \nn \eea and two Higgs
doublets remain as zero modes.

\subsection{Gauge symmetry breaking by brane fields}

Further breaking of Pati-Salam to the Standard Model gauge group
occurs at the symmetry breaking fixed point $y = \pi R / 2$. On
the Pati-Salam brane $\chi^c \equiv {\bf (\bar{4}, 1, \bar 2)}$,
which might come from a ${\bf 16}$ of $SO(10)$, can get a vev
along the $\bar \nu_{sterile}$ direction. The low energy gauge
group then becomes that of the Standard Model (see Figure 1).  In
order to preserve supersymmetry we require both $\chi^c + \bar
\chi^c$ to get vevs.

The breaking scale of the Pati-Salam gauge group is not
significantly constrained by proton decay and can be as low as
$\sim 10$ TeV\cite{Deo:pw,Weinberg:1980bf}. However, it is well
known that supersymmetric Pati-Salam, broken at an intermediate
scale $\sim 10^{12}$ GeV, cannot achieve unification without
additional charged fields\cite{Deshpande:1996jv}.  An acceptable
prediction for gauge coupling unification can be obtained however
in a minimal supersymmetric Pati-Salam gauge group when PS is
broken at the unification scale (for recent analyses of
supersymmetric Pati-Salam, see \cite{Allanach:1995sj} -
\cite{King:2000vp}). Thus we take the brane breaking scale to be
the cutoff scale, i.e the highest scale we can imagine, and at the
same time, the true unification scale. It is also very natural to
expect the symmetry breaking scale to be of order the cutoff
scale.

Suppose the brane breaking scale is the cutoff scale $M_*$. The
breaking of the gauge symmetry by the brane field $\chi^c$ (and
$\bar \chi^c$ with $\la \bar \chi^c \ra = \la \chi^c \ra$) gives a
mass to the gauge fields, localized on the
brane\cite{Nomura:2001mf}. \bea {\cal L} & \subset & \left[ \delta
(y-\frac{\pi R}{2})+ \delta (y-\frac{3\pi R}{2}) \right] \la
\chi^c \ra^2 A^{\hat{a}}_{\mu} A^{\hat{a} \mu}, \eea where
$A^{\hat{a}}_{\mu}$ are the even modes in PS/SM. The wave
functions of the form \bea  A^{\hat{a}}_{\mu n} (y) & = & N_n \cos
(M_n y) \nn \eea with $N_n$ a normalization constant satisfy \bea
-\partial_y^2 A^{\hat{a}}_{\mu} + \left[ \delta (y-\frac{\pi
R}{2})+\delta (y-\frac{3\pi R}{2}) \right] g_5^2 \la \chi^c \ra^2
A^{\hat{a}}_{\mu} = m^2 A^{\hat{a}}_{\mu} . \eea The eigenvalue
$M_n$ is determined by the jumping condition at $y=\pi R/2$ (and
also at $y=3\pi R/2$) \bea 2M_n \sin (\f{M_n \pi R}{2}) + g_5^2
\la\chi^c\ra^2 \cos (\f{M_n \pi R}{2}) = 0. \eea
The eigenvalues for $M_n \ll M_* \sim \la\chi^c\ra$ are easy to
find and are given by\footnote{We could also relax the constraint
$M_n \ll M_*$. However, it will become clear later that, only this
limit gives acceptable gauge coupling unification.}
\bea M_n & \simeq & \f{(2n+1 - \f{n+1/2}{N} \zeta)}{R} \nn \\&
\simeq & \f{(2n+1 - \f{n}{N} \zeta)}{R}, \eea with $M_*/M_c = 2N$
and \bea \zeta & = & \f{4M_*}{\pi g_5^2 \la \chi^c \ra^2}.
\label{zeta} \eea The approximation is valid up to $n \sim N$ if $
\zeta < 1 $. Naive dimensional analysis with the strong coupling
assumption \cite{Chacko:1999hg} gives $g_5^2 \simeq
\f{24\pi^3}{M_*}$ and $\la\chi^c\ra \simeq \f{M_*}{4 \pi}$.
Therefore, we have $\zeta = \zeta_0 \simeq 8/3\pi^2 \simeq 0.27 <
1$.

Note, although in the above discussion we have just considered the
corrections to the KK spectrum of the gauge bosons, it should be
clear that the brane Higgs mechanism preserves N = 1
supersymmetry. As a result the bulk gauge bosons and gauginos, the
chiral $\Sigma$ field and the brane Higgs fields combine to form
degenerate massive N = 1 supermultiplets. In addition, we have
worked in a unitary gauge for the gauge sector, so that the
goldstone mode is absorbed. Moreover, we assume that all other
physical Higgs degrees of freedom acquire mass at the cutoff
scale.  Hence they do not affect the running of the gauge
couplings.   Finally, the brane breaking does not affect the
spectrum of bulk Higgs fields.

\begin{figure}
\begin{center}
\begin{picture}(400,480)(-20,-40)
  \Line(10,0)(400,0)
  \LongArrow(10,0)(10,420)
  \Text(0,0)[r]{$0$}
  \GBox(10,29)(400,30){0.4}
  \GBox(10,87)(400,90){0.4}
  \GBox(10,145)(400,150){0.4}
  \GBox(10,203)(400,210){0.4}
  \GBox(10,261)(400,270){0.4}
  \GBox(10,319)(400,330){0.4}
  \GBox(10,377)(400,390){0.4}
  \GBox(10,30)(400,60){0.7}
  \GBox(10,90)(400,120){0.7}
  \GBox(10,150)(400,180){0.7}
  \GBox(10,210)(400,240){0.7}
  \GBox(10,270)(400,300){0.7}
  \GBox(10,330)(400,360){0.7}
  \GBox(10,390)(400,420){0.7}
  \Line(8,30)(12,30) \Text(6,30)[r]{$1/R$} \DashLine(10,30)(400,30){1}
  \Line(8,60)(12,60) \Text(6,60)[r]{$2/R$}\DashLine(10,60)(400,60){1}
  \Line(8,90)(12,90) \Text(6,90)[r]{$3/R$}\DashLine(10,90)(400,90){1}
  \Line(8,120)(12,120)\DashLine(10,120)(400,120){1}
  \Line(8,150)(12,150)\DashLine(10,150)(400,150){1}
  \Line(8,180)(12,180)\DashLine(10,180)(400,180){1}
  \Line(8,210)(12,210)\DashLine(10,210)(400,210){1}
  \Line(8,240)(12,240)\DashLine(10,240)(400,240){1}
  \Line(8,270)(12,270)\DashLine(10,270)(400,270){1}
  \Line(8,300)(12,300)\DashLine(10,300)(400,300){1}
  \Line(8,330)(12,330)\DashLine(10,330)(400,330){1}
  \Line(8,360)(12,360)\DashLine(10,360)(400,360){1}
  \Line(8,390)(12,390)\DashLine(10,390)(400,390){1} \Text(6,420)[r]{$M_*$}
  \GCirc(60,0){3}{1}
  \Vertex(60,60){3}
  \Vertex(60,120){3}
  \Vertex(60,180){3}
  \Vertex(60,240){3}
  \Vertex(60,300){3}
  \Vertex(60,360){3}
  \DashLine(10,29)(400,29){2}
  \Vertex(180,29){3}\DashLine(10,87)(400,87){7}
  \Vertex(180,87){3}\DashLine(10,145)(400,145){7}
  \Vertex(180,145){3}\DashLine(10,203)(400,203){7}
  \Vertex(180,203){3}\DashLine(10,261)(400,261){7}
  \Vertex(180,261){3}\DashLine(10,319)(400,319){7}
  \Vertex(180,319){3}\DashLine(10,377)(400,377){7}
  \Vertex(180,377){3}
  \Vertex(300,30){3}
  \Vertex(300,90){3}
  \Vertex(300,150){3}
  \Vertex(300,210){3}
  \Vertex(300,270){3}
  \Vertex(300,330){3}
  \Vertex(300,390){3}
  \Text(60,-20)[b]{$(V,\Phi)_{SM}$}
  \Text(180,-20)[b]{$(V,\Phi)_{PS/SM}$}
  \Text(300,-20)[b]{$(V,\Phi)_{SO(10)/PS}$}
\end{picture}
\caption{Mass spectrum for Kaluza-Klein modes of vector multiplets
of the Standard Model and Pati-Salam/Standard Model and
$SO(10)$/Pati-Salam. A white blob is the zero mode of the standard
model vector multiplets and is $V_{SM}$ rather than
$(V,\Phi)_{SM}$. The regions are divided into three parts. White
regions are the `region I', dark gray regions are the `region II'
and light gray regions are the `region III'. The figure shows that
the region II appears near the cutoff scale, and in most ranges
the full interval from $M_c$ to $M_*$ is governed by the region I
and the region III.}
\end{center}
\end{figure}
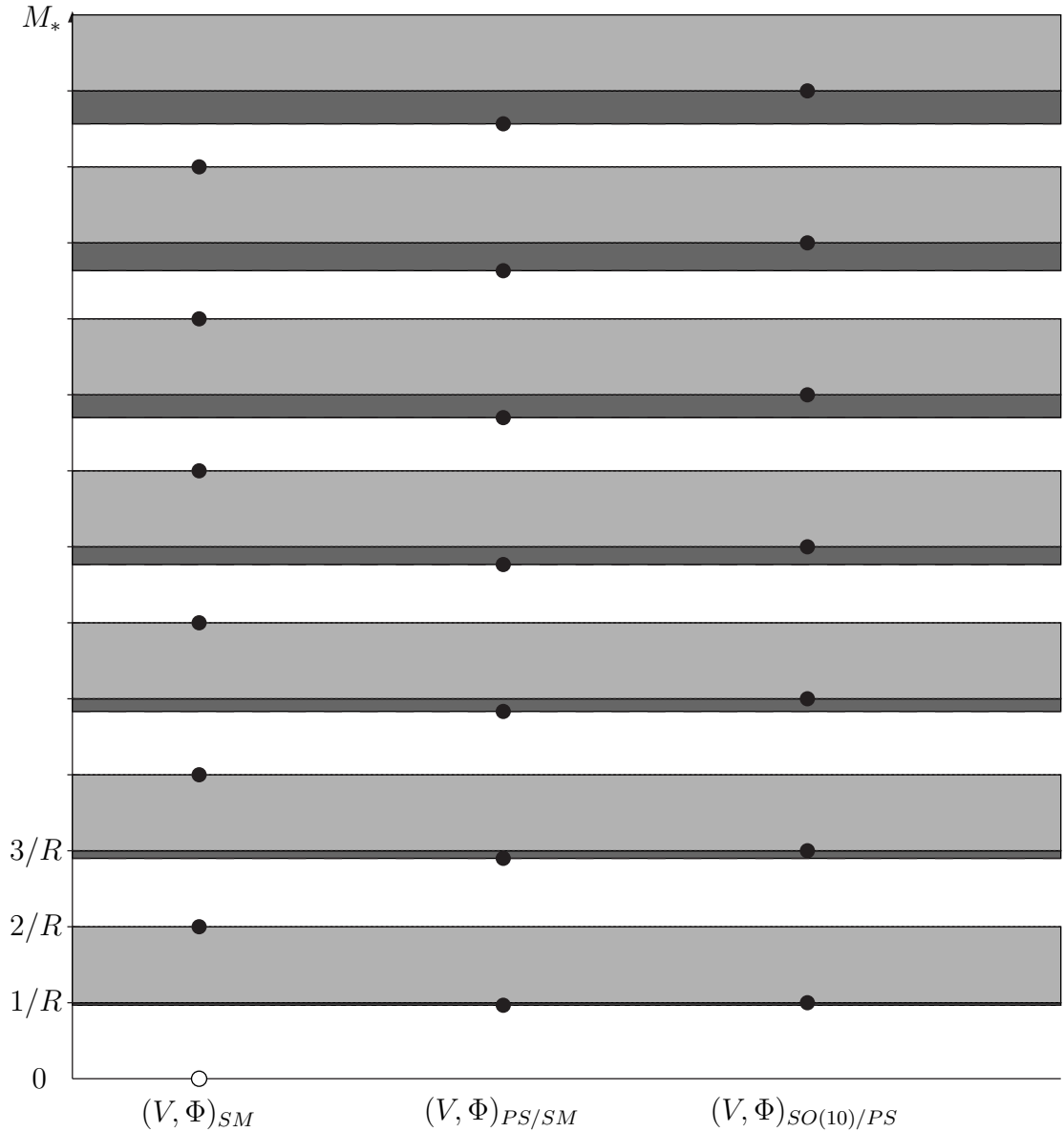

\section{Running of gauge couplings in hybrid model}

The model considered here contains two different types of symmetry
breaking. One is the orbifold breaking of $SO(10)$ to Pati-Salam
and the other is the Higgs breaking on the brane of Pati-Salam to
the Standard Model. The running of gauge couplings with either
orbifold breaking\cite{Hall:2001pg,Hebecker:2001wq} or with brane
breaking\cite{Nomura:2001mf} has been studied extensively.
However, the combined case has not been considered before. Here we
derive the formulae for gauge coupling running in the hybrid model
with both orbifold and brane breaking.

Though our concern is $SO(10)$ breaking to the Standard Model via
Pati-Salam, we use a more general notation in which the original
bulk gauge group $F (\equiv SO(10))$ is broken down to $G (\equiv
PS)$ by orbifolding and the brane breaking leaves $H (\equiv SM)$
only at low energy: \bea F \supset G \supset H. \nn \eea The 5D
vector multiplet $V_F$ is divided into $V_{G++}$ and $V_{F/G+-}$.
The zero modes of vector multiplets in $F/G$ are lifted up by
nontrivial twisting of the boundary condition in the orbifolding
procedure. Again $G$ is divided into $H$ and $G/H$ and only
$V_{H++}$ remain massless. ($\Sigma$ has the opposite $Z_2$ and
$Z_2^{\prime}$ parities.)

A five dimensional gauge theory is nonrenormalizeable and gets a
large correction at the cutoff. The cutoff dependent
uncontrollable corrections to the gauge couplings come in an $F$
invariant way and affect the absolute values of the gauge
couplings at the unification.  Nonetheless, in the case of
orbifold and/or brane breaking we can reliably calculate the
running of $1/\alpha_3 - 1/\alpha_1$ and $1/\alpha_2 -
1/\alpha_1$.  Thus, we can address the question of unification by
examining the differences of the couplings even though we can not
exactly predict the absolute value of the gauge couplings at the
GUT scale.   The one loop renormalization of gauge couplings is
given by \bea \f{2\pi}{\alpha_i (\mu)} & = & \f{2\pi}{\alpha
(M_*)} + b_i \log ( \frac{M_c}{\mu}) + \Delta_i . \label{eq:rge}
\eea  The beta function coefficient is \bea b_i = -3 C_2(SU(i)) +
T(R_f), \nn \eea with $C_2$, the quadratic Casimir of the $SU(i)$
gauge group,  $T$, the Dynkin index for the chiral multiplet in
the representation $R_f$ (normalized to 1/2 for the fundamental
representation of $SU(i)$) and, for the abelian group $U(1)_Y$, $
b_1 = \frac{3}{5} \sum_i Y_i^2$. Also, $\Delta_i$ are the
threshold corrections due to all the KK modes from $M_c$ to $M_*$
and can be expressed as $\Delta_i = b^{eff}_i \log
(\frac{M_*}{M_c})$.

We use the following notation for the one loop beta function
coefficients $ b^A_i $ in which $A$ represent a group $F,G,H$ or a
coset space $F/G,F/H,G/H$ and $i$ runs from 1 to 3 representing
the Standard Model gauge group. \bea b^A = (b^A_1,b^A_2,b^A_3),
\,\,\, \Delta = (\Delta_1,\Delta_2,\Delta_3) \nn \eea will be used
for three component vectors of the beta function coefficients and
the threshold corrections respectively. Then for the 4D MSSM, we
have $b=(33/5,1,-3)$ for $U(1), \ SU(2)$ and $SU(3)$ respectively.
Also $b = b_v + b_h + b_m$ where $b_v = (0,-6,-9)$, $b_h =
(3/5,1,0)$ and $b_m = (6,6,6)$ are the $v =$ vector, $h =$ Higgs
and $m =$ matter contributions. It is clear that $b_m$ is not
relevant for discussing unification since it is common to three
gauge couplings. Thus when considering unification, where only the
difference of the three gauge couplings is important, we consider
$b = b_{v} + b_{h} = (3/5,-5,-9) \equiv (0,-28/5,-48/5)$, i.e. we
subtract any contribution proportional to the vector (1,1,1)  (see
appendix \ref{beta}).

Now it is important to understand the spectrum for the symmetry
breaking to get the threshold corrections. $F/G$ is only affected
by the orbifolding and $G/H$ is only affected by the brane
breaking. Therefore there are three different relevant energy
regions. First of all, there is a region I in which only the
Standard Model gauge group $H$ is involved. Then a region II
where, in addition to $H$, $G/H$ begins to appear and we have $H
\oplus G/H = G$.  Finally, a region III, where we have $H \oplus
G/H \oplus F/G = F$. The one loop beta function coefficients
differ for the three different regions.

\subsection{Gauge symmetry breaking by orbifolding}

Let us begin with the orbifold breaking (no brane breaking). We
have $F$ broken down to $G$; thus at this stage we do not have
mass splitting between $H$ and $G/H$. The orbifold breaking splits
Higgs doublets from Higgs triplet. Therefore, up to the
compactification scale, \bea & b^I_{\rm gauge} & = b^G(V), \nn \\
& b^I_{\rm Higgs} &= b (H_2). \eea Just above the compactification
scale $1/R$, $F/G$ vector multiplets and a pair of two Higgs
triplets appear and we have \bea &b^{III}_{\rm gauge} & = b^G(V) +
b^{F/G}(V) + b^{F/G}(\Sigma), \nn \\
&b^{III}_{\rm Higgs} & = b (H_2) + 2 b (H_3). \nn \eea (Note, in
the next section we show region I breaks up into two regions I and
II.) Since we know \bea
b^F(V) & = &  b^G(V) + b^{F/G}(V) \nn \\
b^F(\Sigma) & = & b^{G}(\Sigma) + b^{F/G}(\Sigma), \nn \eea and
$b^F$ is common to three Standard Model gauge group, as long as
the difference of the couplings is concerned, we can subtract the
common contribution. Then \bea &b^{III}_{\rm gauge} & =
-b^{G}(\Sigma). \nn \eea Also from the appendix \ref{beta}, $b
(H_3) = - b (H_2)$ and we have \bea &b^{III}_{\rm Higgs} & = - b
(H_2). \nn \eea

Again above $\f{2}{R}$, Kaluza-Klein states of $G$ vector
multiplets and a pair of two Higgs doublets appear and \bea
&b^{I}_{\rm gauge} & = b^G(V) + b^{G}(\Sigma) + b^{III} = b^G(V), \nn \\
&b^{I}_{\rm Higgs} & = -b (H_2) + 2 b (H_2) = b (H_2).
 \nn \eea The region $[\f{(2n-1)}{R},\f{2n}{R}]$ is
III and the region $[\f{2n}{R},\f{(2n+1)}{R}]$ is I for $1 \le n
\le N$ with \bea M_* & \sim & \f{(2N+1)}{R}. \nn \eea

\subsection{Inclusion of brane breaking}

The spectrum of the gauge sector lifted by the brane breaking is
given in section 2.3. The brane breaking repels the wave functions
of the bulk even modes. Therefore, for $M_{n} \ll \la \chi^c \ra$
(or the brane breaking scale), the effective parity of the bulk
vector fields changes from $++$ to $+-$ . (Supersymmetry also
requires the parity of the adjoint scalars to change from $--$ to
$-+$ .) Thus the vector multiplets in $H$ are not affected, but
those in $G/H$ are lifted up and are nearly degenerate with $F/G$.
There is no effect on the odd parity modes since the wave function
already vanishes on the brane.   For $M_{n} \gg \la \chi^c \ra$,
the even mode spectrum is recovered, i.e. the mass shift due to
brane breaking is negligible. Finally, for $M_{n} \sim \la \chi^c
\ra$ we are in an intermediate region where the mass of the vector
multiplets in $G/H$ are in between those of $H$ and $F/G$ at each
KK level. As a result of these mass shifts we show that the region
I in the previous section separates into two parts, namely, the
region I and II.  Region III, however, is not affected.

The spectrum in the Higgs sector is not affected by the brane
breaking.  Hence, for the Higgs sector there is no distinction
between regions I and II. Therefore the result is the same as in
the previous section \bea
b^I_{\rm Higgs} & = & b^{II}_{\rm Higgs} = b (H_2), \nn \\
b^{III}_{\rm Higgs} & = & - b (H_2). \nn \eea

In order the get the beta function coefficients of the gauge
sector, let us now consider the three regions in more detail (see
Figure 3). Before the brane breaking is taken into account, region
I has both the vector multiplets of $H$ and $G/H$. Now it only has
those of $H$. Then there appears a new region, called region II,
for which $V_H$, $V_{G/H}$ and $\Sigma_{G/H}$ are included. For
$M_n \ll \la \chi^c \ra$, $G/H$ is nearly degenerate with $F/G$
and effectively there is no region II. However, for $ M_n \sim \la
\chi^c \ra$, region I is $[\f{2n}{R},\f{(2n+1/2)}{R}]$ and region
II is $[\f{(2n+1/2)}{R},\f{(2n+1)}{R}]$.  Now in region I, \bea
b^{I}_{\rm gauge} & = & b^H(V), \nn \eea valid up to nearly $1/R$.
Just before $1/R$, there is a region II at which $G/H$ is
incorporated. \bea b^{II}_{\rm gauge} & = & b^H(V) + b^{G/H}(V) +
b^{G/H}(\Sigma) \nn \\
& = & b^G (V) + b^G(\Sigma) - b^H(\Sigma), \nn \eea and is valid
up to $1/R$. Finally, in region III, the beta function coefficient
is the same as before. \bea b^{III}_{\rm gauge} & = & b^H(V) +
b^{G/H}(V) + b^{G/H}(\Sigma) + b^{F/G}(V)
+ b^{F/G}(\Sigma) \nn \\
& = & -b^{H}(\Sigma). \nn \eea In summary, region III is given by
$[\f{(2n-1)}{R},\f{2n}{R}]$,  region I by
$[\f{2n}{R},\f{(2n+1)-\zeta \f{n}{N}}{R}]$, and region II by
$[\f{(2n+1)-\zeta \f{n}{N}}{R},\f{(2n+1)}{R}]$.
%

The gauge and Higgs sectors give $b^M = b^M_{\rm gauge} + b^M_{\rm
Higgs}$ for $M=I,II,III$. The final expression for the threshold
correction is \bea \Delta \equiv \ b^{eff} \log (\f{M_*}{M_c}) & =
& b^I A_I  + b^{II} A_{II} + b^{III} A_{III} , \label{eq:delta1}
\eea with \bea A_I & = & \sum_{n=1}^{N-1} \log \left[ \frac{2n+1 -
\frac{n}{N} \zeta}{2n} \right] , \\
A_{II} & = & \sum_{n=1}^{N-1} \log \left[ \frac{2n+1}{2n+1
- \frac{n}{N} \zeta } \right] , \\
A_{III} & = & \sum_{n=1}^N \log \left[ \frac{2n}{2n-1} \right].
\eea $A_I$, $A_{II}$ and $A_{III}$ are the partial sums of logs
for the corresponding regions and $A_I+A_{II}+A_{III} = \log 2N =
\log (M_*/M_c)$.

Using the approximation $ \log (1+x) = x + \cdots$, we have \bea
A_I & = & \sum_{n=1}^N \left( \frac{1}{2n} - \frac{\zeta}{2N}
+ \cdots \right)  \\
& = & \frac{1}{2} \log 2N - \frac{1}{2} \log (\frac{\pi}{2}) -
\frac{\zeta}{2} + {\cal O}(\f{1}{N}) ,\\ A_{II} & = &
\frac{\zeta}{2} + {\cal O}(\f{1}{N}) , \\ A_{III} & = &
\frac{1}{2} \log 2N + \frac{1}{2} \log (\frac{\pi}{2}). \eea

Putting $A_{M}$ for $M=I,II,III$ with neglecting ${\cal O}
(\f{1}{N})$, we obtain \bea \Delta & = & \f{1}{2}(b^{III}+b^I)
\log (\f{M_*}{M_c}) + \f{1}{2}
(b^{III} - b^I) \log (\f{\pi}{2}) \nn \\
&&+ \f{1}{2} (b^{II}-b^I) \zeta. \label{master} \eea

The threshold correction is $\Delta = \Delta_{\rm gauge} +
\Delta_{\rm Higgs}$. For the gauge contributions, using the fact
that $b^A (\Sigma) = - b^A (V)/3$
and the results for $b^M_{\rm gauge}$ with $M=I,II,III$, we get
\bea {\Delta}_{\rm gauge} & = & \f{2}{3} b^H (V) \log
(\f{M_*}{M_c}) - \f{1}{3} b^H (V) \log
(\f{\pi}{2}) \nn \\
&&+ \f{1}{3} (b^G (V) -b^H (V) ) \zeta. \label{gauge} \eea
 For the
Higgs contributions, we have $b^I_{\rm Higgs} = b^{II}_{\rm Higgs}
= b (H_2)$ and $b^{III}_{\rm Higgs} = -b (H_2)$. Putting it to the
Eqn. (\ref{master}) gives only a tiny finite contribution \bea
{\Delta}_{\rm Higgs}  & = & - b (H_2) \log (\f{\pi}{2}).
\label{Higgs} \eea

In Eqns. (\ref{eq:delta1},~\ref{master}-\ref{Higgs}) we give the
general expression for the threshold corrections at $M_c$ to gauge
coupling running in 5D with orbifold and brane breaking. It is
interesting that the first two terms in Eqn. (\ref{gauge}) depend
only on $H$($\equiv$ the SM gauge group) and are independent of
$F$ and $G$.  In addition, the contribution of the Higgs sector
(Eqn. (\ref{Higgs})) is tiny.  Therefore, when the brane breaking
is extremely large ($\zeta \ll 1$), the running of the difference
of the gauge couplings is determined solely by the gauge sector of
the SM gauge group.

\section{Checking gauge coupling unification}

Using Eqns. (\ref{master} - \ref{Higgs}) and the results of
appendix \ref{beta}, we get \bea \Delta & = &(0, -4 \log
\f{M_*}{M_c} + \frac{8}{5} \log(\frac{\pi}{2}) + \frac{14
\zeta}{5}, -6 \log \f{M_*}{M_c} + \frac{18}{5}
\log(\frac{\pi}{2})+ \frac{9 \zeta}{5} ) . \eea


Let us now make contact with the low energy data.   This will
allow us to fix the scales $M_c$ and $M_*$ (or $N$).  At tree
level in 4D GUTs we have  $\alpha_i, \ i= 1,2,3$ all equal at the
GUT scale.   However, with the precision electroweak data, SUSY
GUTs are now being tested using two loop renormalization group
running from $M_{GUT}$ to $M_Z$ with one loop threshold
corrections at both the GUT and weak scales.   The weak scale
corrections depend on sparticle masses.  However, even with this
weak scale correction, one still needs GUT threshold corrections
in order to precisely fit the low energy data.   Once GUT
threshold corrections are included, the definition of the scale
$M_{GUT}$ becomes ambiguous. We define the scale $M_{GUT}$ as the
scale where $\alpha_1(M_{GUT}) = \alpha_2(M_{GUT}) = \tilde
\alpha_{GUT}$ in the equivalent 4D theory (meaning ignoring the
effect of KK modes). However, at $M_{GUT}$, we have
$\alpha_3(M_{GUT}) = \tilde \alpha_{GUT} ( 1 + \epsilon_3 )$ where
$\epsilon_3 \equiv (\alpha_3(M_{GUT}) - \tilde
\alpha_{GUT})/\tilde \alpha_{GUT} \approx -0.04$ is necessary to
fit the data (see for example, \cite{Blazek:1999ue,Raby:er} and
Altarelli et al. \cite{Babu:1998wi}). In a 4D SUSY GUT, the value
of $\epsilon_3$ depends logarithmically on the masses of states at
the GUT scale. This is typically dominated by the states in the
GUT breaking and Higgs sectors. In our 5D theory, these are
replaced by the towers of KK modes. In fact, we can now use the
values of $M_{GUT}, \ \tilde \alpha_{GUT}$ and $\epsilon_3$ to fix
$M_c$ and $M_*$ (see Figure 4).

In this paper we have evaluated the logarithmic threshold
corrections to $\alpha_i$ at the compactification scale $M_c$ due
to all the KK modes.   We can now use the low energy data to find
the values of the gauge couplings at $M_c$.  Let $\delta_i(\mu)
\equiv 2 \pi (\f{1}{\alpha_i(\mu)} - \f{1}{\alpha_1(\mu)})$ for $i
= 2,3$.  In the MSSM, only the massless modes contribute to the
running from $M_c$ to $M_{GUT}$, i.e. \bea \f{2\pi}{\alpha_i
(\mu)} & = & \f{2\pi}{\alpha (M_{GUT})} + b_i \log (
\frac{M_{GUT}}{\mu} ) \label{eq:rgemssm} \eea however the
difference $\delta_2(M_c)$ is fixed by the KK modes. Using \bea 0
\equiv \delta_2(M_{GUT}) = \delta_2(M_c) + (b_2 - b_1)
\log(\f{M_{GUT}}{M_c}) \eea (with only massless modes included in
the running, Eqn. (\ref{eq:rge})) we find \bea -\frac{28}{5} \log
(\f{M_{GUT}}{M_c}) & = & -4 \log (\f{M_*}{M_c}) + \frac{14
\zeta}{5}+ \frac{8}{5} \log(\frac{\pi}{2}) . \label{eq:delta2}
\eea

Then using \bea \delta_3(M_{GUT}) & = & - 2\pi
(\f{\epsilon_3}{\tilde \alpha_{GUT}})  \label{eq:delta3} \\ & = &
2\pi \left( \frac{1}{\alpha_3(M_{c})} -
\frac{1}{\alpha_{1}(M_{c})} \right) - (b_3 - b_1)
\log(\f{M_{GUT}}{M_c}) \nn \\ & = &- 6 \log (\f{M_*}{M_c}) +
\frac{18}{5} \log(\frac{\pi}{2}) + \frac{9 \zeta}{5} +
\frac{48}{5} \log (\f{M_{GUT}}{M_c}) \nn \\& \sim & \frac{2\pi
\times 0.04}{\tilde \alpha_{GUT}} \sim 6.0, \eea  with $\tilde
\alpha_{GUT} = 24$.

\begin{figure}
\begin{center}
\begin{picture}(320,220)(-20,-20)
\LongArrow(0,0)(0,220) \Line(0,200)(290,200)
\DashLine(200,178)(250,210){3} \DashLine(200,190)(250,200){3}
\Line(0,50)(200,178) \Line(0,150)(200,190) \Line(200,190)(280,200)
\Line(200,178)(280,200) \Text(-10,45)[b]{$\delta_3$}
\Text(-10,145)[b]{$\delta_2$} \LongArrow(-5,0)(320,0)
\Text(320,-15)[b]{$\mu$} \Line(0,-4)(0,6) \Line(200,-4)(200,6)
\Line(240,-4)(240,6) \Line(280,-4)(280,6)
\Text(0,-18)[b]{$M_{SUSY}$} \Text(200,-18)[b]{$M_c$}
\Text(250,-18)[b]{$M_{\rm GUT}$} \Text(280,-18)[b]{$M_*$}
\LongArrow(250,205)(250,210) \LongArrow(250,205)(250,200)
\Text(253,205)[l]{$\epsilon_3$}
\end{picture}
\caption{Differential running of $\delta_2 = 2\pi
(\frac{1}{\alpha_2} -\frac{1}{\alpha_1})$ and $\delta_3 = 2\pi
(\frac{1}{\alpha_3} -\frac{1}{\alpha_1})$.}
\end{center}
\end{figure}
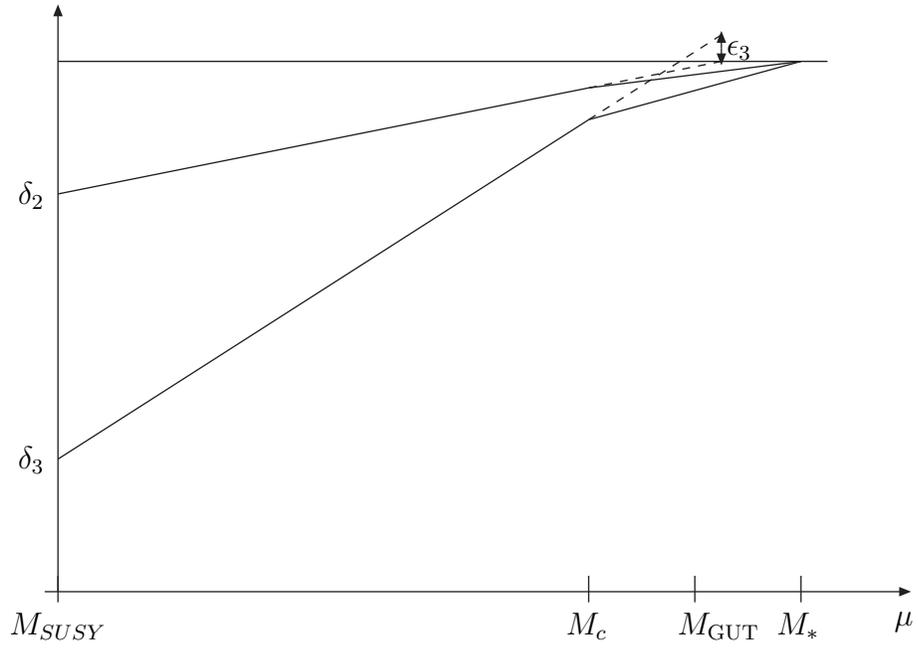

Combining Eqns. (\ref{eq:delta2}, \ref{eq:delta3}), we find \bea
\log (\f{M_*}{M_c}) & \simeq & 6.8 + 3.0 \zeta, \nn \\ \log
(\f{M_{GUT}}{M_c} ) & \simeq & 4.8 + 1.7 \zeta, \nn \eea and using
$M_{GUT} =  3 \times 10^{16} \GeV$ we finally obtain, \bea
M_c & = & 1.5 \times 10^{14} \GeV , \\
M_* & = & 3.2 \times 10^{17} \GeV. \nn \eea for $\zeta = \zeta_0 =
\frac{8}{3\pi^2} = 0.27$ (see discussion after Eqn. \ref{zeta}).

Hence,  the three 4D SUSY GUT parameters $M_{GUT}, \ \tilde
\alpha_{GUT}, \ \epsilon_3$ are replaced by the 5D SUSY GUT
parameters  $M_c, \ M_*, \alpha_3(M_c)$.  Note, the
compactification scale $M_c$ is significantly less than $M_{GUT}$.
It is the compactification scale which determines the rate for
proton decay on the $SO(10)$ brane.  We discuss this further in
section \ref{sec:pdecay}.

Note also our results are not very sensitive to the value of the
arbitrary parameter $\zeta$.  For example, if $\zeta$ is
$\frac{1}{2} \zeta_0$, \bea
M_c & = & 1.9 \times 10^{14} \GeV , \nn \\
M_* & = & 2.7 \times 10^{17} \GeV. \nn \eea and if $\zeta$ is
$2\zeta_0$, \bea
M_c & = & 1.0 \times 10^{14} \GeV , \nn \\
M_* & = & 4.7 \times 10^{17} \GeV. \nn \eea

If there are more fields than $\chi^c$ and $\bar \chi^c$ which
acquire vevs, $\zeta$ becomes smaller and will tend to go to zero,
and in that limit we obtain, \bea
M_c & = & 2.4 \times 10^{14} \GeV , \label{eq:zeta0}\\
M_* & = & 2.2 \times 10^{17} \GeV, \nn \eea which is the same as
that of the 5D $SU(5)$ orbifold model.

In the limit $\zeta =0$ (infinitely large brane vev), Eqn.
(\ref{gauge}) and (\ref{Higgs}) can be simplified by redefining
the compactification scale (or the matching scale) as \bea \hat
M_c & \equiv & \f{2}{\pi} M_c . \eea The finite threshold
correction with $\log (\f{\pi}{2})$ is then absorbed and \bea \hat
{\Delta}_{\rm gauge} & =
& \f{2}{3} b^{SM}_i (V) \log (\f{M_*}{\hat M_c}) \\
\hat {\Delta}_{Higgs}  & = & 0,
 \eea with the threshold correction $\hat \Delta$ defined at $\hat
 M_c$ as
\bea \f{2\pi}{\alpha_i (\mu)} & = & \f{2\pi}{\alpha (M_*)} +
b^{MSSM}_i \log ( \frac{\hat M_c}{\mu}) + \hat \Delta_i \\
& = & \f{2\pi}{\alpha (M_*)} + b^{MSSM}_i \log ( \frac{\hat
M_c}{\mu}) + \f{2}{3} b^{SM}_i (V) \log (\f{M_*}{\hat M_c}).
\label{5d} \eea Here $b^{MSSM}$ includes the gauge sector and the
Higgs sector together and $b^{SM}(V)$ includes the gauge sector
only. See appendix \ref{beta} where our notation is defined. The
running equation is very simple and permits us to directly compare
with well known 4D SUSY GUTs. In a 4D MSSM, the running equation
is given as \bea \f{2\pi}{\alpha_i (\mu)} & = & \f{2\pi}{\alpha
(M_{GUT})} + b^{MSSM}_i \log ( \frac{m_{H_3}}{\mu}) + b^{SM}_i (V)
\log (\f{M_{GUT}}{m_{H_3}}) . \label{4d} \eea with lower mass for
the triplet Higgs, $m_{H_3} < M_{X,Y} = M_{GUT}$. If we neglect
the constraints for the triplet Higgs mass coming from the proton
lifetime, we can achieve unification by adjusting the color
triplet Higgs mass $m_{H_3} = 2 \times 10^{14} \GeV$
\cite{Raby:er}.

Comparing Eqn. (\ref{5d}) and (\ref{4d}), we observe that if \bea
 \hat M_c =  m_{H_3}, \,\,\,\,\,
 \f{M_*}{\hat M_c}  =
(\f{M_{GUT}}{m_{H_3}})^{\f{3}{2}}, \nn \eea the same unification
is achieved here. Therefore we get $\hat M_c = 2 \times 10^{14}
\GeV$ and $M_* = 3.7 \times 10^{17} \GeV$ by using $M_{GUT} = 3
\times 10^{16} \GeV$.  Note this is in reasonable agreement with
the results in Eqn. (\ref{eq:zeta0}), since this is derived with a
one loop analysis with exact unification at $M_{GUT}$ whereas the
value of $M_{GUT} = 3 \times 10^{16} \GeV$ is derived from a 2
loop analysis including one loop threshold corrections at the GUT
scale.

However, gauge coupling unification would break down if we had
assumed the GUT breaking vev on the brane was of order $M_c$
instead of $M_*$.   In this case the PS/SM states would have
adversely affected the good GUT results.

Finally, before we continue we should consider whether such a
large ratio  $M_*/M_c \sim 10^3$ makes sense. Even though the
differences of gauge couplings run logarithmically, the individual
couplings satisfy a power law running above $M_c$. This is a
consequence of the higher dimensional physics or equivalently it
can be obtained as a result of the changing of the beta function
coefficients when crossing the thresholds of the tower of KK
modes. The evolution of the individual couplings satisfies the
approximate renormalization equation ( \cite{Dienes:2002bg} or Kim
et al. \cite{Hebecker:2001wq}) \bea \f{2 \pi}{\alpha_i (M_*)} &
\approx & \f{2 \pi}{\alpha_i (M_c)} + \Delta_i -  \f{\tilde
b_i}{2} [\f{M_*}{M_c} - 1] \label{eq:powerlaw} \eea where the
coefficients $\tilde b_i$ only receive contributions from bulk
modes.  Since however the KK modes come in complete $SO(10)$
multiplets we have $\tilde b_i \equiv \tilde b$ for all $i$.   As
argued in \cite{Dienes:2002bg}, if $ \tilde b < 0$, all gauge
couplings approach an ultra violet fixed point, $\alpha_* =
\alpha_i (M_*) \approx - 4 \pi M_c/\tilde b M_* $ for all $i$. In
the case of $SO(10)$ we have $\tilde b = \tilde b_{gauge} + \tilde
b_{matter}$ with $\tilde b_{gauge} = - 16$ and $ \tilde b_{matter}
= 2 n_{10} + 4 n_{16}$ where $n_{10} \ ( n_{16} )$ equals the
number of hypermultiplets in the $10 \ ( 16 )$ dimensional
representation. Note, for $n_{10} = 1, \  n_{16} = 3$ (i.e. one
Higgs hypermultiplet and three 16s in the bulk), $\tilde b < 0$
and there is a possible UV fixed point.  Hence a large ratio
$M_*/M_c \sim 10^3$ would not be impossible.\footnote{In the field
theory with a UV fixed point, there is no need for the cutoff
$M_*$, since the field theory is renormalizable at the fixed
point. Nevertheless, our field theory does not include gravity,
hence we must identify $M_*$ with the physical scale ($M_{Planck}$
and/or $M_{string}$) at which gravitational physics is included.}
We note that although this result Eqn. (\ref{eq:powerlaw}) is
perturbative and only valid in a specific regularization
scheme\cite{Dienes:2002bg}, there are indications that
non-perturbative UV fixed points may exist in 5D non-abelian gauge
theories\cite{Seiberg:1996bd}.  For example, in Intriligator et
al.\cite{Seiberg:1996bd} it was shown that for $SO(10)$ a
non-perturbative fixed point exists for any $n_{10} \leq 6$ and
$n_{16} \leq 2$.  The model considered in this paper clearly
satisfies this condition.   Note, however, some recent $SU(5)$
GUTs in 5D do not.  For example, in the complete $SU(5)$ model of
Ref. \cite{Hall:2002ci}, the third family is placed on the
symmetric $SU(5)$ brane, in order to preserve bottom-tau Yukawa
unification. However in order to prevent rapid proton decay, the
first generation and parts of the second are put in the bulk.   As
a result, $\tilde b
> 0$ and a large ratio $M_*/M_c$, although necessary for gauge
coupling unification, is unlikely.  The non-perturbative condition
for an UV fixed point in this case\cite{Seiberg:1996bd} is more
severe.  It allows a maximum of one $\bf 10$ and three ${\bf \bar
5}$s
 in the bulk. Note, since in $SU(5)$ it takes two $ ({\bf 10} + {\bf
\bar 5})$s in the bulk to obtain one massless family, it would not
be possible to put any families in the bulk.

\section{Proton decay \label{sec:pdecay}}
The dangerous baryon and lepton number violating dimension five
operators are forbidden by a $U(1)_R$ symmetry\cite{Hall:2001pg}.
This is a crucial difference between 5D orbifold GUTs and 4D SUSY
GUTs. Also, a discrete subgroup of $U(1)_R$, $R$-parity, forbids
the dimension four baryon and lepton number violating operators.
Of course, it would be necessary to construct a complete $SO(10)$
theory, including fermion masses and mixings, in order to be
certain that this $U(1)_R$ symmetry can be preserved.

The size of any possible dimension six operators depends crucially
on the placement of matter fields. There are three possibilities
for the matter field configurations.
\begin{itemize}
\item They can be five dimensional fields: They appear as zero
modes of hypermultiplets in the bulk. \item They can live on the
$SO(10)$ brane. \item They can live on the PS brane.
\end{itemize}
When all matter fields live on the $SO(10)$ preserving brane,
there are dimension six operators which mediate proton decay
obtained by integrating out the X and Y gauge bosons (belonging to
the $(6,2,2)$ representation of PS). These operators come with a
factor $1/M_c^2$ and the current bounds on the proton lifetime
give a direct lower bound on the compactification scale of $5
\times 10^{15} \GeV$ \cite{Murayama:2001ur}.   Since this bound is
violated in our model, only the third family can be placed on the
$SO(10)$ brane.

However in the other cases, for matter fields in the bulk or on
the PS brane, there is no direct process of mediation for the
proton to decay via the X and Y gauge bosons, unless as discussed
in \cite{Hebecker:2002rc}, there are new dimension five operators
generated on the PS brane of the form \bea  {\cal L}_\psi =
\f{c_{ij}}{M_*} \delta(y^\prime) \int d^2 \theta d^2 \bar \theta \
\psi^{c \dagger}_j \ (\nabla_5 e^{2 V}) \ \psi_i  + h.c.
\label{eq:operator}  \eea where $\nabla_5 =
\partial_5 + \Sigma$, the constant $c$ is determined by strong
interactions at the cutoff and the fields $\psi_i, \ \psi^c_j$ are
$i$th and $j$th generation matter fields in PS representations.
This operator allows the X and Y gauge bosons to couple to the
matter multiplets. Upon integrating out the X and Y gauge bosons,
it generates the effective dimension 6 operator of the form \bea
{\cal O} \sim \f{b \ c_{ij} c_{kl} \ g_4^2}{M_c \ M_*} \int d^2
\theta d^2 \bar \theta \ \sum_{\hat a} (\psi^{c \dagger}_i T^{\hat
a} \psi_j)(\psi^{c \dagger}_k T^{\hat a} \psi_l) \eea with $b$
another unknown strong interaction parameter and $i,j,k$ and $l$
are flavor indices running from 1 to 3. It results in a proton
lifetime given by\cite{Hebecker:2002rc} \bea 1/\Gamma(p
\rightarrow \pi^0 e^+) = 3.5 \times 10^{34} {\rm years} \times
\f{1}{b^2 c_{11}^4} (\f{M_c}{10^{15} {\rm GeV}})^2
(\f{M_*}{10^{17} {\rm GeV}})^2 \nn \eea

Taking values of $b \sim c_{11} \sim 1$ and $M_c, \ M_*$
consistent with gauge coupling unification we find a proton
lifetime \bea 1/\Gamma(p \rightarrow \pi^0 e^+) \sim 7 \times
10^{33 \pm 2} {\rm years}. \eea   The $\pm 2$ in the exponent
characterizes the uncertainties in the unknown strong interaction
parameters $b, c$ and the precise value of the cutoff and
compactification scales $M_*, \ M_c$.

\section{Conclusion}
We have discussed gauge coupling unification in a 5D $SO(10)$
model with hybrid (orbifold/Higgs) breaking of the gauge symmetry.
The three 4D SUSY GUT parameters $M_{GUT}, \ \tilde \alpha_{GUT},
\ \epsilon_3$ are replaced by the 5D SUSY GUT parameters  $M_c, \
M_*, \alpha_3(M_c)$ with $M_c \sim 1.5 \times 10^{14} \GeV$ and
$M_* \sim 3 \times 10^{17} \GeV$.  Note, the compactification
scale $M_c$ is significantly less than $M_{GUT}$.  It is the
compactification scale which determines the rate for proton decay
on the $SO(10)$ brane.  However in this theory, all the families
can be placed on the PS brane.  This makes any calculation of
proton decay extremely model dependent.

There are many virtues in this 5D $SO(10)$ model. Firstly, it
retains an explanation of the SM charges of quarks and leptons and
$U(1)_Y$ charge quantization by having the Pati-Salam gauge group
at the orbifold fixed point. Secondly, both the orbifold and brane
breaking generate only logarithmic corrections to the differences
of the three gauge couplings. This allow us to maintain a
calculable and predictive unification of gauge couplings,
competitive with 4D GUTs. Thirdly, doublet-triplet splitting is
nicely explained and all unwanted extra states become heavy by
boundary conditions. Fourthly, the Pati-Salam symmetry is
sufficient to preserve many of the good GUT symmetry relations for
quark and lepton masses.  Finally, the conditions for a
non-perturbative UV fixed point is satisfied in our model. Hence
the large ratio $M_*/M_c \sim 10^3$, necessary to preserve gauge
coupling unification in all higher dimensional GUTs, is plausible.
It remains to construct a complete 5D $SO(10)$ model with
predictive fermion masses and mixing angles. See, for example,
\cite{Albright:2002pt} for recent work in this direction.

\acknowledgments   Partial support for this work came from DOE
grant\# DOE/ER/01545-839.


\appendix
\section*{Appendix}
\section{$SO(10)$ breaking to the Standard Model via Pati-Salam}

In the appendix, we summarize how the representation of $SO(10)$
is decomposed into that of the Pati-Salam gauge group. \bea SO(10)
& \rightarrow & G_{PS} = SU(4)_C \times SU(2)_L \times SU(2)_R \nn
\\
 {\bf 45} & \rightarrow & ({\bf 15},{\bf 1},{\bf 1})
\oplus ({\bf 1},{\bf 3},{\bf 1}) \oplus ({\bf 1},{\bf 1},{\bf 3})
\oplus ({\bf 6},{\bf 2},{\bf 2}), \\
{\bf 10} & \rightarrow & ({\bf 6},{\bf 1},{\bf 1}) \oplus
({\bf 1},{\bf 2},{\bf 2}), \\
{\bf 16} & \rightarrow & ({\bf 4},{\bf 2},{\bf 1}) \oplus ({\bf
\bar{4}},{\bf 1},{\bf 2}). \eea Now $SU(4)_C$ is broken down to
$SU(3)_C \times U(1)_{B-L}$.
\bea
SU(4)_C & \rightarrow & SU(3)_C \times U(1)_{B-L} \nn \\
{\bf 4} & \rightarrow & {\bf 3}_{1/3} \oplus {\bf 1}_{-1} , \\
{\bf 15} & \rightarrow & {\bf 8}_0 \oplus {\bf 3}_{4/3}
\oplus {\bf \bar{3}}_{-4/3} \oplus {\bf 1}_0 , \\
{\bf 10} & \rightarrow & {\bf 6}_{2/3}
\oplus {\bf 3}_{-2/3} \oplus {\bf 1}_{-2} , \\
{\bf 6} & \rightarrow & {\bf 3}_{-2/3} \oplus {\bf \bar{3}}_{2/3}.
\eea At the same time, $SU(2)_R$ is broken to $U(1)_{T_3}$ and the
hypercharge is given as $Y  =  T_{3R} + \frac{1}{2}(B-L)$ .
Combining these together, we get \bea
& {\bf 45} & : SU(3)_C \times SU(2)_L \times U(1)_Y \nn \\
& ({\bf 15},{\bf 1},{\bf 1})_+ & \rightarrow ({\bf 8},{\bf 1})_0
\oplus ({\bf 3},{\bf 1})_{2/3} \oplus ({\bf \bar{3}},{\bf
1})_{-2/3} \oplus ({\bf
1},{\bf 1})_0, \\
 & ({\bf 1},{\bf 3},{\bf 1})_+ & \rightarrow ({\bf 1},{\bf 3})_0 , \\
 & ({\bf 1},{\bf 1},{\bf 3})_+ & \rightarrow ({\bf 1},{\bf 1})_1
 \oplus ({\bf 1},{\bf 1})_0 \oplus ({\bf 1},{\bf 1})_{-1} , \\
 & ({\bf 6},{\bf 2},{\bf 2})_- & \rightarrow ({\bf 3},{\bf 2})_{1/6}
 \oplus ({\bf 3},{\bf 2})_{-5/6} \oplus
({\bf \bar{3}},{\bf 2})_{5/6} \oplus ({\bf \bar{3}},{\bf 2})_{1/6} ,\\
 & {\bf 10} & : \nn \\
 & ({\bf 6},{\bf 1},{\bf 1})_- & \rightarrow ({\bf 3},{\bf 1})_{-1/3}
 \oplus ({\bf \bar{3}},{\bf 1})_{1/3} , \\
 & ({\bf 1},{\bf 2},{\bf 2})_+ & \rightarrow ({\bf 1},{\bf 2})_{1/2}
 \oplus ({\bf 1},{\bf 2})_{-1/2} , \\
 & {\bf 16} & : \nn \\
 & ({\bf 4},{\bf 2},{\bf 1})_{-i} & \rightarrow ({\bf 3},{\bf 2})_{1/6}
 \oplus ({\bf 1},{\bf 2})_{-1/2} , \\
 & ({\bf \bar{4}},{\bf 1},{\bf 2})_{i} & \rightarrow
 ({\bf \bar{3}},{\bf 1})_{1/3} \oplus
({\bf \bar{3}},{\bf 1})_{-2/3} \oplus ({\bf 1},{\bf 1})_1 \oplus
({\bf 1},{\bf 1})_0 , \eea in which the subscript ($+,-$) or
($i,-i$) of the Pati-Salam representations denotes the $B-L$
parity $P^{\prime} = \exp{(-i \f{3\pi}{2} (B-L))}$ which is used
to break $SO(10)$ to Pati-Salam gauge group.

\section{1 loop beta function coefficients}
\label{beta} Since we deal with differential running, we subtract
contributions proportional to the vector (1,1,1) and make $b^A_1 =
0$.
\begin{itemize}
\item MSSM gauge sector
\begin{enumerate}
\item MSSM vector multiplet : \bea b^{SM} (V) & = & (0,-6,-9).
\eea \item SUSY Pati-Salam vector multiplet : \bea b^{PS} (V) & =
& (0,12/5,-18/5). \eea \item SUSY $SO(10)$ vector multiplet: \bea
b^{SO(10)} (V) & = & (0,0,0). \eea
\end{enumerate}
\item MSSM Higgs sector = Two Higgs doublets : \bea b (H_2) & = &
(0,2/5,-3/5). \eea \item Two Higgs triplets : \bea b (H_3) & = & -
b (H_2) = (0,-2/5,3/5). \eea
\end{itemize}
Therefore, we get \bea b^{MSSM} & = & (0,-28/5,-48/5). \eea

One loop beta function coefficients of the hybrid model when
$SO(10)$ is broken down to the Standard Model via Pati-Salam gauge
group are given below.
\begin{itemize}
\item{Region I}
  \bea
 & b^I_{\rm gauge} & =  b^{SM} (V), \nn \\
 & b^I_{\rm Higgs} & =  b (H_2) , \nn \\
 & b^I & =  b^I_{\rm gauge} +
b^I_{\rm Higgs} \nn \\
&&= (0,-6,-9) + (0,2/5,-3/5) = (0,-28/5,-48/5) .\nn \eea
\item{Region II}
 \bea
& b^{II}_{\rm gauge} & =  b^{PS} (V) + b^{PS/SM} (\Sigma), \nn \\
&  b^{II}_{\rm Higgs} & =  b (H_2) , \nn \\
&  b^{II} & =  b^{II}_{\rm gauge} + b^{II}_{\rm Higgs} \nn \\
&&= (0,-2/5,-27/5) + (0,2/5,-3/5) = (0,0,-6) .\nn \eea
\item{Region III}  \bea & b^{III}_{\rm gauge} & =  b^{SO(10)}
(V) + b^{SO(10)/SM} (\Sigma) = -b^{SM} (\Sigma), \nn \\
& b^{III}_{\rm Higgs} & =  b (H_3) =  - b (H_2) , \nn \\
& b^{III} & =  b^{III}_{\rm gauge} + b^{III}_{\rm Higgs} \nn \\
&&= (0,-2,-3) + (0,-2/5,3/5) = (0,-12/5,-12/5).\nn \eea
\end{itemize}


\end{document}